\documentclass[oldversion,rnote, preprint]{aa} 

\usepackage{graphicx}
\usepackage{txfonts}
\begin{document}
   \title{1.6\,GHz VLBI Observations of SN\,1979C: almost-free expansion}
   \titlerunning{Almost-free expansion of SN\,1979C.}

   \author{J.M. Marcaide\inst{1}
          \and
          I. Mart\'i-Vidal\inst{1,2}
          \and
          M.A. Perez-Torres\inst{3}
          \and
          A. Alberdi\inst{3}
          \and
          J.C. Guirado\inst{1}
          \and
          E. Ros\inst{2,1}
          \and
          K.W. Weiler\inst{4}
          }

   \institute{Dpt. Astronomia i Astrof\'isica, Universitat de Val\`encia,
              C/ Dr. Moliner 50, E-46100 Burjassot (Spain)\\
              \email{J.M.Marcaide@uv.es}
         \and
             Max-Planck-Institut f\"ur Radioastronomie,
             Auf dem H\"ugel 69, D-53121 Bonn (Germany)
         \and
             Instituto de Astrof\'isica de Andaluc\'ia (CSIC),
             C/ Camino bajo de Hu\'etor 50, E-18008 Granada (Spain)
         \and
             Naval Research Laboratory, Washington D.C., USA
             }

   \date{Submitted to A\&A on 14 May 2009. Accepted on 7 Jul 2009.}

  \abstract
{We report on 1.6\,GHz Very-Long-Baseline-Interferometry (VLBI)
observations of supernova SN\,1979C made on 18 November 2002. We
derive a model-dependent supernova size. We also present a reanalysis
of VLBI observations made by us on June 1999 and by other authors
on February 2005. We conclude that, contrary to our earlier
claim of strong deceleration in the expansion, SN\,1979C has been
undergoing almost-free expansion ($ m = 0.91\pm0.09$; $R \propto
t^m$) for over 25 years.}

   \maketitle
%

\section{Introduction}

Supernova SN\,1979C was discovered in the Virgo Cluster galaxy
M\,100, at a distance of 16.1$\pm$1.3\,Mpc (Ferrarese et al.
\cite{Ferrarese1996}), on 19 April 1979 (Mattei et al.
\cite{Mattei1979}). It reached an absolute magnitude of $\sim-20$,
becoming one of the brightest supernovae ever observed in the optical
band (e.g., Young \& Branch \cite{Young1989}). Radio emission from
SN\,1979C was detected and monitored by Weiler \& Sramek
(\cite{Weiler1980}) at several frequencies. From the analysis of
the radio lightcurves, Weiler et al. (\cite{Weiler1986}) estimated
the explosion date of the supernova to be 4 April 1979, 15 days
before its discovery. The peak flux density observed by Weiler et
al. was $\sim$9.8\,mJy at 20\,cm. This emission level and the
large distance to the host galaxy, M\,100, made supernova
SN\,1979C the most luminous radio supernova (RSN) at its time.

Several VLBI observations of this supernova have been made since
year 1982 through year 2005 at 5, 2.3, and 1.6\,GHz (see Bartel et
al. \cite{Bartel1985}, \cite{Bartel2003}, \cite{Bartel2008};
Bartel \cite{Bartel1991}; and Marcaide et al.
\cite{Marcaide2002}). From these observations, the expansion curve
of the supernova was determined by two research groups. Each one
arrived at very different conclusions. Bartel (\cite{Bartel1991})
found that SN\,1979C was freely expanding over the first 7 years
after explosion. Later, Marcaide et al. (\cite{Marcaide2002}),
using the early expansion curve published by Bartel et al.
(\cite{Bartel1985}) together with optical-line data and new VLBI
observations made in year 1999, claimed a ``strong deceleration''
in the supernova expansion, starting $\sim$6 years after
explosion. In year 2002, we performed new VLBI observations 
of SN1979C in order to confirm this strong deceleration. Later,
Bartel \& Bietenholz (\cite{Bartel2003}) 
reported on VLBI observations made in years 1990, 1996, and 2001, 
and claimed a practically free expansion over almost two decades.
These results were in clear conflict with those published by
Marcaide et al. (\cite{Marcaide2002}).

In the next section, we describe the details of these new VLBI
observations of SN\,1979C and the calibration scheme used. In
Sect. \ref{III}, we present the results obtained from the data
analysis. In Sect. \ref{ComparSec}, we describe the results
obtained after a reanalysis of the observations reported in
Marcaide et al. (\cite{Marcaide2002}). In Sect. \ref{IV}, we
present a reanalysis of the observations recently reported by
Bartel \& Bietenholz (\cite{Bartel2008}). In Sect. \ref{V} we
report on the expansion curve of SN\,1979C. Finally, in Sect.
\ref{VI} we summarize our conclusions.

\section{Observations and Data Reduction}

Our 1.6\,GHz observations were made on 18 November 2002, from
2:20\,UT to 21:00\,UT. The participating stations were: the
complete Very Long Baseline Array (VLBA; 10 antennas, 25\,m
diameter each, spread over the USA), Green Bank (100\,m diameter,
West Virginia, USA), Effelsberg (100\,m diameter, Germany),
Robledo (70\,m, Spain), Westerbork (phased array of size
equivalent to a 93\,m diameter antenna, The Netherlands), and
Arecibo (300\,m diameter, Puerto Rico). The recording rate was set
to 256\,Mbps, 2-bit sampling, obtaining a total
bandwidth of 64\,MHz for all stations. The data were
cross-correlated at the Array Operations Center of the National
Radio Astronomy Observatory (NRAO), in Socorro, New Mexico, USA.

The observations were scheduled in phase-reference mode. Each
cycle time was divided into a $\sim$5\,min long scan of the
calibrator source PKS\,B1157+156 and a $\sim$15\,min long scan of
SN\,1979C. The slewing time of the slowest antenna between sources
lasted typically about 1.5\,min. Every 2$-$3 duty cycles, a
5\,min scan of a secondary calibrator, TXS\,1214+161, was
added. The source 3C\,274 was also observed as a fringe finder at
the beginning of each VLBI tape.

Given the long duration of the experiment and the different
latitudes of the stations, it was not possible to assign only one
reference antenna for the calibration of the whole data set.
Therefore, the observations were divided into two parts for their
calibration, the first one (from 2:20~UT to 8:20~UT) was calibrated 
using Robledo as reference antenna, and the second one (from 8:20~UT 
onwards) was calibrated using North Liberty as reference antenna.

The cross-correlated data were imported into the NRAO program {\sc
aips} for calibration. The phases of the 8 different IFs (of 8 MHz
width each) were manually aligned by fringe-fitting scans of
3C\,274 and applying the resulting antenna phases and delays to
all observations. Afterwards, the amplitude calibration was
performed using system temperatures registered at all the stations
and gain curves for the antennae. Once a hybrid image of
PKS\,B1157+156 was obtained, a standard phase-reference
calibration of SN\,1979C was performed taking into account the
structure of PKS\,B1157+156. Finally, the visibility amplitude
calibration for SN\,1979C was refined by performing an amplitude
self-calibration of the PKS\,B1157+156 visibilities and
interpolating the resulting gains to the SN\,1979C scans. The
calibrated data were then exported from {\sc aips} into the
Caltech software {\sc difmap} (Shepherd, Pearson \& Taylor
\cite{Shepherd1995}) for further reduction and imaging.

\section{Results}
\label{III}

\begin{figure}
\centering
\includegraphics[width=8cm]{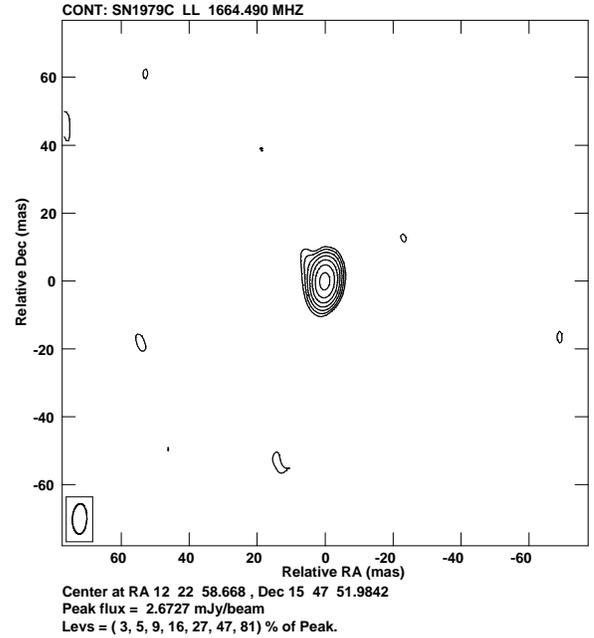}
\caption{CLEAN image of SN\,1979C obtained from the observations made on 
18 November 2002. The Full Width at Half Maximum (FWHM) of the convolving 
beam is shown at the bottom-left corner (a $8.9\times4.2$~mas beam at a 
position angle of $-2.6^{\circ}$).}
\label{SN1979C-MAP}
\end{figure}

\begin{figure}
\centering
\includegraphics[width=9cm]{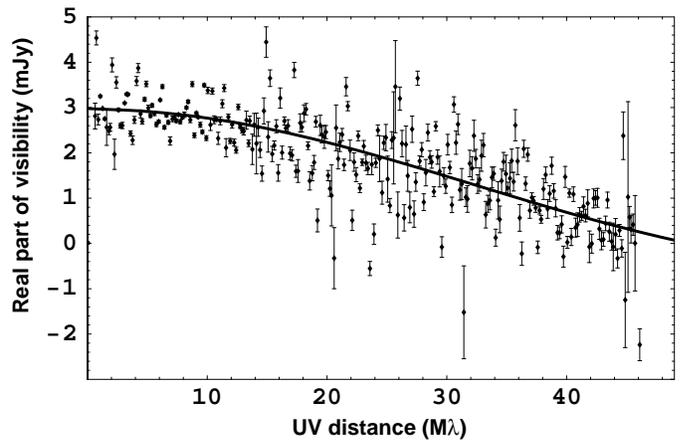}
\caption{Real part of the visibilities taken at epoch 18 November 2002 as 
a function of distance in Fourier space. The visibilities have been 
averaged into 300 radial bins for obtaining a clear plot. The solid line 
corresponds to a 30\%-wide spherical shell model fitted to the 
visibilities (see text and Table \ref{MODELFIT}).}
\label{SN1979C-VISIB}
\end{figure}

We Fourier-inverted the calibrated visibilities of SN\,1979C in
{\sc difmap}. Using a CLEAN deconvolution, we obtained the image
shown in Fig. \ref{SN1979C-MAP}. The total flux density of the
image is 2.96\,mJy and the root-mean-square (rms) of the image
residuals 0.07\,mJy/beam. This phase-referenced image of SN\,1979C
is point-like. However, a modulation of the real
part of the visibilities as a function of uv-distance is readily
observable in the data (see Fig. \ref{SN1979C-VISIB}). That is,
the source structure is partially resolved by the interferometer.
We fitted two models (a uniform sphere and a 30\%-wide spherical
shell) to the visibilities and obtained the corresponding
estimates of the source size. The $\chi^2$-minimization was
performed using the Levenberg-Marquardt algorithm (e.g., Gill \&
Murray 1978), as implemented in the {\em Mathematica 5.0} package
(Wolfram 2003). The parameter uncertainties were computed from the
diagonal elements of the post-fit covariance matrix. In Table
\ref{MODELFIT}, we summarize the results of the fits performed.
The fitted radii are incompatible with those reported in Marcaide
et al. (\cite{Marcaide2002}).

\section{Reanalysis of Our Previous Observations}
\label{ComparSec}

We reanalyzed the observations reported in Marcaide et al.
(\cite{Marcaide2002}). In Table \ref{MODELFIT}, we show the
results of such a reanalysis, which differ substantially 
from those reported in the original paper. 
Thus, our conclusion is that something was not correctly reported in 
Marcaide et al. (\cite{Marcaide2002}). In fact, the model used in the fit 
reported by Marcaide et al. (\cite{Marcaide2002}) was not really a
30\%-wide shell, as said, but a uniform disc with its inner
region removed (i.e., the flux density from radii smaller than
0.7 times the disc radius had been removed). The use of such a
modified disc model results in a radius estimate of
78\% of the radius estimate of the correct 30\%-wide
spherical-shell model. Actually, Marcaide et al.
(\cite{Marcaide2002}) also reported the sizes corresponding 
to an optically thick source (uniform disc) and a ring. These 
two size estimates were correct, and therefore inconsistent with
the size reported for the optically-thin 30\%-wide shell model.
However, this miscalculation went unchecked and the 
inappropriate size estimate was used to compare with the 
optical data and early results by other authors, thus leading to 
an inaccurate conclusion about the expansion of the source.

The deceleration parameter derived by Marcaide et al.
(\cite{Marcaide2002}) from optical-line velocities was obtained
assuming that the supernova had been expanding with a constant
velocity during the first 5 years after explosion.
Basically, the authors assigned the mean optical-line velocity
measured during the first 6 weeks after explosion to the expansion
velocity during the fifth year after explosion. The main argument
for such an assumption was that the ratio of the early 
optical-line velocity and the velocity derived from VLBI at year 5 
after explosion was $\sim$0.7, that expected for a
30\%-wide shell of a supernova (Marcaide et al. \cite{Marcaide2002}, 
 \cite{Marcaide2009}).
However, if the assumption of a free expansion over the first
5 years is relaxed, things look different.

With an assumption of deceleration of the supernova beginning
on day 70 after explosion, the ejecta velocity at year 5
after explosion would be $\sim$25\% lower than the velocity
estimated from a free expansion, but the estimated fractional
shell width would be 0.53; an unrealistically wide shell.
However, this shell width estimate should be interpreted as an
upper bound, given that the optical-line emission may originate in
a cool region behind the shocked ejecta (Chevalier \& Fransson
\cite{Chevalier1994}) and, thus, any comparison between
optical-line velocities and those inferred from VLBI (which are
additionally affected by any uncertainty in the distance
estimate) should not be taken at face value for the derivation of 
the supernova expansion curve.

\begin{figure}
\centering
\includegraphics[width=9cm]{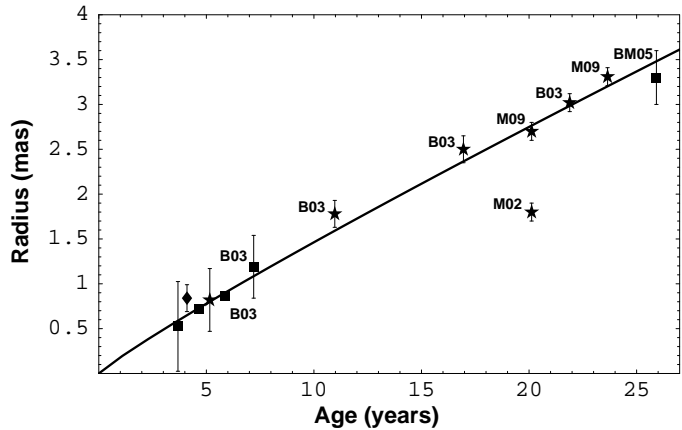}
\caption{Expansion curve of SN\,1979C using all the available VLBI
data. Squares are data at 5\,GHz, diamonds at 2.3\,GHz, and stars
at 1.6\,GHz. All these values (except the one marked with M02)
have been obtained using the same model fitted to the visibilities
(a homogeneous sphere). Epochs until year 5 after explosion are
taken from Bartel et al. (\cite{Bartel1985}), M02 refers to
Marcaide et al. (\cite{Marcaide2002}), B03 refers to Bartel \&
Bietenholz (\cite{Bartel2003}), M09 refers to the observations
here reported (included the re-analysis of the observations
reported in Marcaide et al. \cite{Marcaide2002}), and BM05 refers
to the observations reported in Bartel et al. (\cite{Bartel2008}),
which we have reanalyzed for estimating the source size (see
text). The solid line is a fit with a time power law (i.e., $R =
K\,t^m$, with $m = 0.91\pm0.09$).} 
\label{EXPANSION}
\end{figure}

The fit of a homogeneous sphere to the data was registered in the
research logs, but it was not reported in Marcaide et al.
(\cite{Marcaide2002}). A direct comparison of that size with those
reported by Bartel et al. (\cite{Bartel1985}) is also totally
compatible with an almost free expansion, as is now shown in Fig.
\ref{EXPANSION}. Given that Bartel et al. used a uniform sphere to
fit all their observations, Marcaide et al.
(\cite{Marcaide2002}) should have also used this same model but
instead, and unfortunately, they used a scaling factor, based on
simulations, to {\em convert} the sizes reported by Bartel et al.
(\cite{Bartel1985}) into sizes corresponding to a shell model.

\begin{table}
\begin{minipage}[t]{\columnwidth}
\centering
\caption{Results of modelfitting to the SN\,1979C visibilities.}
\label{MODELFIT}
\begin{tabular}{ c  c | c  c }
\hline\hline
Age & Freq. & \multicolumn{2}{|c}{ Radius (mas)} \\
 (years) & (GHz) & Uniform Sphere & 30\%-Wide Shell \\
\hline
20.12\footnote{Re-analysis of the observations reported in Marcaide et al. (\cite{Marcaide2002}).}
      & 1.6 & 2.70 $\pm$ 0.10 & 2.27 $\pm$ 0.10 \\
23.65\footnote{Observations reported in this paper.}
      & 1.6 & 3.31 $\pm$ 0.10 & 2.81 $\pm$ 0.08 \\
25.92\footnote{Re-analysis of the observations reported in Bartel \& Bietenholz (\cite{Bartel2008}).}
      & 5.0 & 3.30 $\pm$ 0.30 & 2.84 $\pm$ 0.26 \\
\hline
\end{tabular}
\end{minipage}
\end{table}

\section{Reanalysis of Observations by Other Authors}
\label{IV}

\begin{figure}
\centering
\includegraphics[width=8cm]{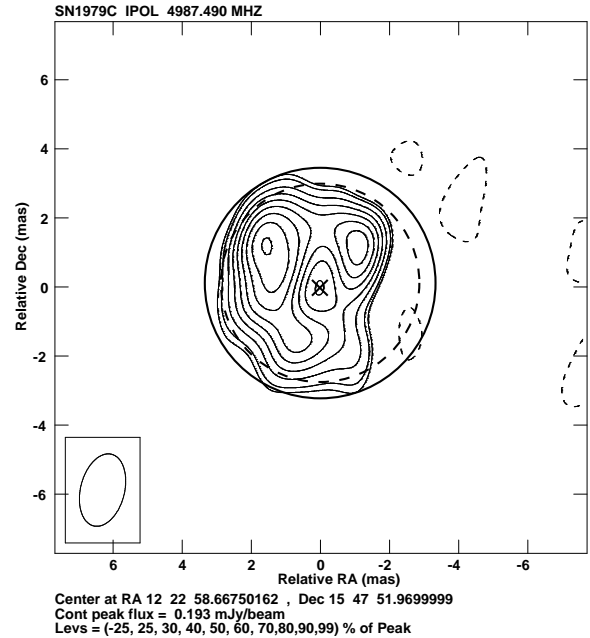}
\caption{CLEAN image obtained after a reanalysis of the
observations reported in Bartel \& Bietenholz (\cite{Bartel2008}). 
The continuous and dashed lines correspond to a sphere and a 30\%-wide 
spherical shell fitted to the visibilities. The center of the models 
has been fixed at the location of the central minimum 
intensity of the image (marked with a cross).} 
\label{BARTEL08}
\end{figure}

Bartel \& Bietenholz (\cite{Bartel2008}) have recently reported on
new VLBI observations of SN\,1979C, made at 5\,GHz on 25 February
2005. From these observations, the authors found a shell-like
structure for SN\,1979C. However, the authors postponed an
estimation of the source size to a future publication. In order to
obtain the most complete expansion curve, we have reanalyzed their
data (already public NRAO archive data), following the same steps
described in Bartel \& Bietenholz (\cite{Bartel2008}), except for
our decision not to phase self-calibrate the data given the high
noise of the visibilities compared to the flux density of the
supernova. Our decision translates into a slightly higher rms 
of the map residuals (20\,$\mu$Jy\,beam$^{-1}$ in our image compared
to the 11\,$\mu$Jy\,beam$^{-1}$ reported by Bartel \& Bietenholz).
Otherwise, our image (shown in Fig. \ref{BARTEL08}) is similar to
that reported in Bartel \& Bietenholz (\cite{Bartel2008}).

In our fits to determine the size of the source, we only used the
visibilities with distances in Fourier space shorter than
50\,M$\lambda$. This way we avoided any possible bias arising from
the use of different shell resolutions between epochs. We centered
the fitting models at the location of the central intensity
minimum of the image (where the shell center is supposed to be). Had we 
set the source position as a free parameter in the fit, it would have
biased the estimated source size, due to the large shell
inhomogeneities. The center of the fitted source would have fallen
close to the image peak and the fitted size would have been,
therefore, mainly related to the size of the main blob of the
image, instead of to the size of the shell.

The CLEAN image of SN\,1979C, together with the fitted sphere and
spherical-shell model, is shown in Fig. \ref{BARTEL08}. The sizes
estimated from the fits to the visibilities are also shown in
Table \ref{MODELFIT}.

\section{The Expansion Curve}
\label{V}

Figure \ref{EXPANSION} shows all the available size
estimates of SN\,1979C over the last 25 years.
Estimates from epochs earlier than 5 years after explosion are
from Bartel et al. (\cite{Bartel1985}), estimate labelled M02 is from 
Marcaide et al. (\cite{Marcaide2002}),
estimates labelled B03 are from Bartel \& Bietenholz (\cite{Bartel2003}),
estimates labelled M09 refer to the observations reported here and to
those wrongly reported as M02 and here reanalyzed, and estimate BM05 refers
to the observations reported in Bartel et al. (\cite{Bartel2008})
and here reanalyzed.  All size estimates (but M02) have been obtained
model fitting a homogeneous sphere to the visibilities as in Bartel et al.
(\cite{Bartel1985}).

Modelling the expansion curve with a standard $\chi^2$
minimization using a time power law (i.e., $R \propto t^m$, see
Chevalier \cite{Chevalier1982}) gives an expansion index $ m =
0.91\pm0.09$, compatible with a free expansion ($m = 1$). Thus, we
conclude that supernova SN\,1979C has been expanding without 
significant
deceleration for more than two decades. Actually, including BM05
in the fit does not change the result at all. At this
epoch the size appears a bit smaller than expected but the
uncertainty in the size determination is large (and it 
could be even larger considering that our criteria 
to center the fitting model may be partially 
inadequate.)

\section{Conclusions}
\label{VI}

We report on 1.6\,GHz VLBI observations of SN\,1979C made on
November 2002. The phase-referenced image of the supernova does
not show a clear structure and has a total flux density of
$2.96\pm0.06$\,mJy. The size
estimates, compared to all the other available VLBI results (but
those reported in Marcaide et al. \cite{Marcaide2002}) are
compatible with a nearly-free expanding supernova for more than
two decades. We reanalyzed the observations reported in Marcaide
et al. (\cite{Marcaide2002}) and found that, for reasons given in
the text, the results and conclusions then published were not
correct.

The expansion curve resulting from the reanalysis of all the VLBI
data (including the observations reported in Bartel \& Bietenholz
\cite{Bartel2008}) results in a new expansion model which is
compatible with a nearly-free expansion for over more than two
decades (expansion index of 0.91$\pm$0.09).

\begin{acknowledgements}

The National Radio Astronomy Observatory is a facility of the
National Science Foundation operated under cooperative agreement
by Associated Universities, Inc. 
The 100-m telescope at Effelsberg is a facility of the MPIfR 
(Max-Planck-Institut f\"ur Radioastronomie).
The Westerbork Synthesis Radio Telescope is operated by the Netherlands 
Institute for Radio Astronomy ASTRON, with support of NWO. 
The Arecibo
Observatory is the principal facility of the National Astronomy and 
Ionosphere Center, which is operated by the Cornell University under 
a cooperative agreement with the National Science Foundation. This 
work has been partially founded by grants AYA2006-14986-CO2-01 and 
AYA2005-08561-C03 of the Spanish DGICYT. KWW wishes to thank the 
Office of Naval Research Laboratories for the 6.1 funding supporting 
this research. IMV is a fellow of the Alexander von Humboldt Foundation. 
\end{acknowledgements}

\end{document}